\documentclass[twocolumn,color,superscriptaddress,psfig,showpacs,amsmath,amssymb,floatfix,fleqn]{revtex4-1}
\usepackage{graphicx}
\usepackage{subfigure}
\usepackage{leftidx}
\usepackage{amsmath}
\usepackage{float}
\usepackage{color}
\usepackage[misc]{ifsym}
\usepackage[colorlinks,
            linkcolor=blue,
            anchorcolor=blue,
            citecolor=blue,
            urlcolor=blue,
            ]{hyperref}

\usepackage{bookmark}

\usepackage{silence}
\DeclareMathOperator{\sech}{sech}

\begin{document}


\title{Impulse control of chaos in the flexible shaft rotating-lifting system of the mono-silicon crystal puller}

\author{Zi-Xuan Zhou}%
\affiliation{
Shaanxi Key Laboratory of Complex System Control and Intelligent Information Processing, Xi'an University of Technology, Xi'an, Shaanxi 710048, People¡¯s Republic of China
}%
\affiliation{%
North Minzu University Library, Yinchuan, Ningxia 750021, People¡¯s Republic of China
}%
\author{Celso Grebogi}
\affiliation{
Shaanxi Key Laboratory of Complex System Control and Intelligent Information Processing, Xi'an University of Technology, Xi'an, Shaanxi 710048, People¡¯s Republic of China
}%
\affiliation{%
Institute for Complex System and Mathematical Biology, University of Aberdeen AB24 3UE, United Kingdom
}%

\author{Hai-Peng Ren}
 \email{renhaipeng@xaut.edu.cn}
\affiliation{
Shaanxi Key Laboratory of Complex System Control and Intelligent Information Processing, Xi'an University of Technology, Xi'an, Shaanxi 710048, People¡¯s Republic of China
}%

%




\date{\today}

\begin{abstract}
Chaos is shown to occur in the flexible shaft rotating-lifting (FSRL) system of the mono-silicon crystal puller. Chaos is, however, harmful for the quality of mono-silicon crystal production. Therefore, it should be suppressed. Many chaos control methods have been proposed theoretically and some have even been used in applications. For a practical plant displaying harmful chaos, engineers from a specified area usually face with the challenge to identifying chaos and to suppressing it by using a proper method. However, despite of the existing methods, chaos control method selection in the FSRL system is not a trivial task. For example, for the OGY method, if one cannot find a practical adjustable parameter, then the OGY method cannot be applied. An impulsive control method is being proposed which is efficiently able to suppress chaos in the FSRL system. The selection of the control parameters is obtained by using the Melnikov method. Simulation results show the correctness of our theoretical analysis and the effectiveness of the proposed chaos control method.

\end{abstract}


\maketitle

\section{Introduction}

As a compact dynamics in nonlinear deterministic system with sensitive dependence on initial conditions, chaos started to attract attention after Lorenz analyzed the chaotic dynamics in a weather prediction process [1]. There were lots of chaotic phenomena reported in different fields, for instance, Refs. [2-5]. On one hand, some chaos properties are beneficial in different engineering fields. For example, the ergodicity of chaos is used for optimization algorithms [6], the broad spectrum property for spread spectrum communication [7], electromagnetic noise reduction [8], liquid mixing [9], road roller [10], and soil compactor [11]. The sensitive dependence on initial condition and on parameter are used in secure communication and encryption [12-15], and the Lyapunov spectrum invariance property is used for improving wireless communication performance [16-18]. On the other hand, the presence of chaos were reported in industrial plants, such as liquid level control system [19], mill roller system [20], and motor drive system [21], all do compromise the performance of the system. In such cases, chaos need to be suppressed. Since the seminal work of Ott-Grebogi-Yorke on controlling chaos [22], many chaos control methods have been proposed. In general, chaos control methods can be divided into two types according to whether the state feedback is being used. Typical feedback methods include linear state feedback [23], nonlinear state feedback [24] and time-delay feedback [25]; typical chaos control methods without state feedback include periodic perturbation [26] and impulse control [27]. Although there exist many chaos control methods with different features, it is not a trivial task to identify a suitable method to eliminate chaos in a special engineering plant. For example, if the plant has not adjustable parameter, then the OGY method cannot be used. Another example is the avoidance of chaos in a permanent magnet synchronous motor (PMSM), reported in [28]. There are only two variables, i.e., quadrature axis and direct axis stator voltages among three state variables, available for manipulation. Therefore, any control method requiring three manipulated variables regulation will be ineffective to apply to the PMSM system[24].

Chaotic dynamics in the FSRL system of the crystal puller was unknown until the work in [29], which reported systematically the nonlinear dynamics in the FSRL system. Due to lack of theoretical insight to the chaotic dynamics in the past, engineers tried conventional methods to avoid, what seemed to be a strange phenomenon, by fine manufacturing and mechanical adjustments of some parts of the device, as well as, avoiding the process parameter causing chaotic motion in the crystal producing procedure. However, nonlinear dynamics is an inherent property of the system, and, as such, researchers in the field has to play an active role. Because of our analysis of the FSRL system [29], we find that it is difficult to control or regulate such a practical system, because no manipulated variable can be used and no parameter can be regulated in real time. To solve this problem, we propose an impulse control method to regulate rotation speed of the system intermittently, so as to impose as little impact as possible on the crystal growing process. The controller parameters selection is derived by using the Melnikov method [30].

The remaining of the paper is organized as follows. In section 2, the complex dynamics of the FSRL system is revisited. In section 3, an impulsive control method is proposed specific for chaos control in the FSRL system, and the parameter selection rule is derived using the Melnikov method. The conclusions is given in section 4.

\begin{figure}

  \centering
  \includegraphics[scale=0.5]{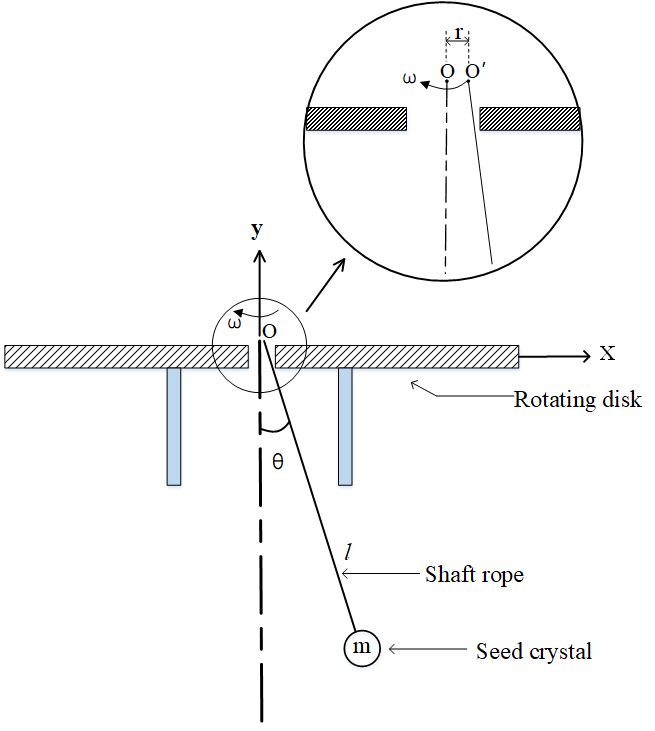}

\caption{The simplified schematic diagram of the FSRL system.}
\end{figure}

\section{System model and the analysis of the controlled system}
A simplified schematic diagram of the FSRL system is given in Fig. 1. During the procedure of the mono-silicon crystal, the mono-silicon seed hanged at the end of the system¡¯s flexible shaft, the FSRL system, which is installed on the rotating disk driven by the rotating motor. In the ideal condition, the rotation of the rotating disk drives the flexible shaft rotation around the center. However, due to the imperfection of the manufacture, the rotating disk might have eccentricity, the actual suspension point $O'$ deviates from
the central axis, the deviation is denoted as $r$. Under the action of rotation, the eccentric effect causes the periodic perturbation $F(t)$.

The modeling details can be found in [29], we learn that the FSRL system can be treated as a pendulum rotating along vertical central axis. Using the Lagrange's equation, the dynamic equation of the FSRL system can be given as:
\begin{equation}%
  \ddot{\theta}=\frac{r}{l}\omega^2\cos(\omega t)+\omega^2\sin\theta\cos\theta-\frac{g}{l}\sin\theta-\frac{\xi}{m}\dot{\theta},  \nonumber
\end{equation}
where $\xi$ is the damping coefficient. Let the above equation convert to dimensionless ones and rewrite it as state space equations:

\begin{align}
  \dot{x_1}&=x_2 \nonumber \\
  \dot{x_2}&=A\Omega^2\cos(\Omega t)+\Omega^2\sin{x_1}\cos{x_1}-\sin{x_1}-cx_2,
  \label{Eqa1}
\end{align}
where $x_1=\theta$, $x_2=\dot\theta$, $A$ and $\Omega$ are the amplitude and frequency of the external excitations respectively, $c$ is the dimensionless damping constant. It should be noticed that the excitation frequency is consistent with the system rotation speed, which is different from the general parametric pendulum.

System (1) can exhibit various dynamical behaviors, including period doubling bifurcation, symmetry-breaking bifurcation, interior crisis and chaotic motion [29]. Figure 2 shows the phase portrait and the corresponding Poincar\'{e} section map for $\Omega=1.2$ with different $A$ and $c$. In Fig. 2a, $A=0.16$, $c=0.1$, in Fig. 2b, $A=0.2$, $c=0.15$, in Fig. 2c, $A=0.27$, $c=0.2$. These complex chaotic behaviors are harmful for the quality of mono-silicon crystal production. Therefore, to get rid of this behaviour, we propose an impulsive control to suppress chaos in the FSRL system.

\begin{figure*}
\centering
\subfigure{
  \includegraphics[scale=0.35]{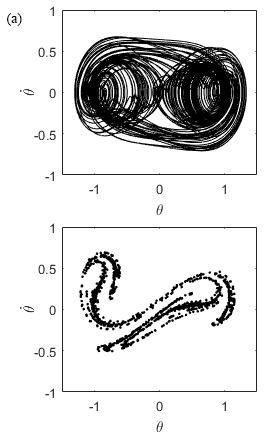}

}
\subfigure{
  \includegraphics[scale=0.35]{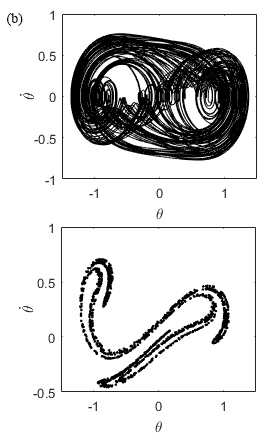}

}
\subfigure{
  \includegraphics[scale=0.35]{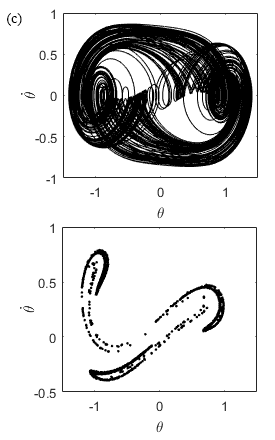}
}

\caption{(a) The phase portrait and the Poincar\'{e} section map for $\Omega=1.2$,$A=0.16$, and $c=0.1$ are shown in the upper panel and in the lower panel, respectively. (b) The phase portrait and the Poincar\'{e} section map for $\Omega=1.2$,$A=0.2$, and $c=0.15$ are shown in the upper panel and in the lower panel, respectively. (c) The phase portrait and the Poincar\'{e} section map for $\Omega=1.2$,$A=0.27$, and $c=0.2$ are shown in the upper panel and in the lower panel, respectively.}
\end{figure*}

Considering the working principle of FRSL system, we apply a periodic impulse to $\Omega$, the equation of the controlled system is then given as follow:

\begin{align}
  \dot{x_1} &=x_2 \nonumber \\
  \dot{x_2} &=A(P(t))^2\cos((P(t))+(P(t))^2\sin{x_1}\cos{x_1}  \nonumber\\
  & -\sin{x_1}-cx_2, \nonumber\\
   P(t) &=\Omega+u(t)
\end{align}
where $u(t)$ is defined as an impulse function with period $T=2\pi/\Omega$, which is expressed as:
\begin{equation}
  u(t)=	\sum_{n=0}^\infty h(t-nT),
\end{equation}
with

\begin{equation} h(t)=
  \begin{cases}
    \kappa,  & -\Delta<t<\Delta \\
    0, & \mbox{otherwise},
  \end{cases}
\end{equation}
$2\Delta$ being the impulse duration and $\kappa$ is the impulse amplitude. Since the impulse perturbation acts on the system rotation speed, we defined a positive impulse as speed increasing and a negative impulse as speed decreasing. We argued in the next section that the chaotic behavior of the system can be suppressed by a proper design of the above impulse control.

\begin{figure*}
  \centering
\subfigure[]{
  \includegraphics[scale=0.5]{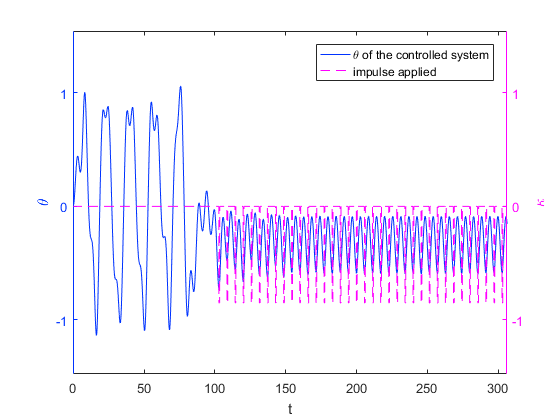}
}
\subfigure[]{
  \includegraphics[scale=0.5]{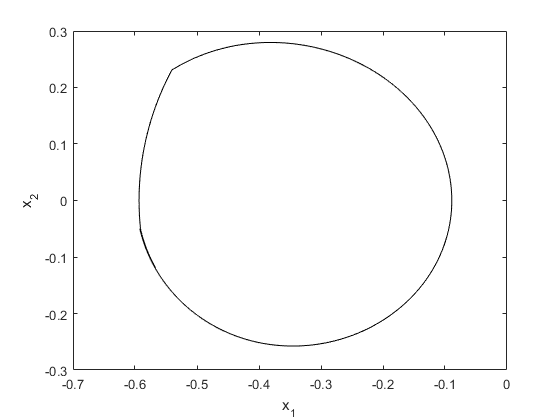}
}
\caption{(a) The state variable $x_1$ is plotted with parameters $\Delta=0.54$ and $\kappa=-0.85$ using blue solid line, magenta dashed line indicates the applied impulse. (b) The corresponding steady state phase portraits after impulse control is activated.}
\end{figure*}

Figure 3a shows state $x_1$ of system (2) and the impulse applied to the system where $\Omega=1.1$, $c=0.1$, and $A=0.2$, and the impulse control is activated at t=100s. The state variable $x_1$ is plotted by a blue solid line, the impulse applied is plotted by a magenta dashed line. Figure 3b shows the steady state phase portrait of the system after the impulse control gose into effect. It can be observed that the chaotic motion in the system is suppressed and is replaced by a periodic motion after the proposed impulse control is imposed.

\section{Impulse control parameter design using the Melnikov method}
In this section, an impulse control method is proposed for suppressing chaos in the FRSL system. Simulations are performed to validate the effectiveness of the proposed method.

In the following, we present the parameter selection procedure for chaos suppression using impulse control.

Since $\Delta$ is small, using $cos(\Omega t)$ as an approximation of $cos(P(t))$, the controlled system equation (2) can be rewritten as:

\begin{align}
  \dot{x_1}&=x_2 \nonumber \\
  \dot{x_2}&=A(\Omega+u(t))^2\cos(\Omega t)+(\Omega+u(t))^2\sin{x_1}\cos{x_1} \nonumber  \\
  & -\sin{x_1}-cx_2,
\end{align}

\textit{Theorem 1.} For system (2), if the impulse duration $\Delta$ and the impulse amplitude $\kappa$ satisfy the inequality (6), the chaotic behavior is eliminated.

\begin{equation}
  \Delta<\dfrac{\dfrac{2c}{A}\times(\dfrac{\ln(\sqrt{\alpha^2-1}-\alpha)}{\sqrt{\alpha^2-1}}+\alpha)-\pi\Omega^2\Phi(\Omega)}{(2\Omega\kappa+\kappa^2)(\Phi(\Omega)+\Psi(n))},
\end{equation}
where
\begin{equation}
  \Phi(\Omega)=\sin [\dfrac{\Omega}{\alpha}\sinh^{-1}(\alpha)]\times\sech (\dfrac{\Omega\pi}{2\alpha}),
\end{equation}
and
\begin{align}%
  \Psi(n) &=\sum_{n=1}^N [(n-1)\sin \dfrac{\Omega}{\alpha}\sinh^{-1}(\alpha)\times\sech (\dfrac{\Omega\pi}{2\alpha}(n-1) \nonumber\\
  & -(1+n)\sin \dfrac{\Omega}{\alpha}\sinh^{-1}(\alpha)\times\sech (\dfrac{\Omega\pi}{2\alpha}(1+n)].
\end{align}

\begin{figure}

  \centering
  \includegraphics[scale=0.5]{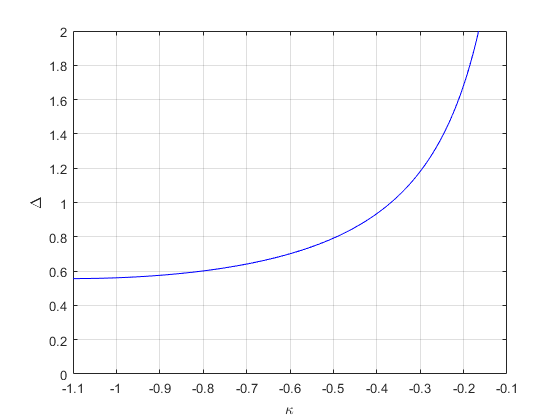}

\caption{The pairs of $(\Delta, \kappa)$ above the curve indicated can be used to suppress chaotic motion.}
\end{figure}

\textit{Proof:} Let $\Omega t=\tau$; system (2) can be written as:
\begin{equation}
\nonumber
  \dot{\boldsymbol{x}}=f(\boldsymbol{x})+\dfrac{1}{\Omega}g(\boldsymbol{x},\tau),
\end{equation}
where
\begin{equation}
\nonumber
    \boldsymbol{x}=
  \begin{pmatrix}   
    x_1 \\  
    x_2\\  
  \end{pmatrix}
  ,
\end{equation}
\begin{equation}
\nonumber
  f(\boldsymbol{x})=
  \begin{pmatrix}   
    x_2/\Omega \\  
    ((\Omega+u(\tau))^2\sin x_1\cos x_1-\sin x_1)/\Omega  
  \end{pmatrix}
  ,
\end{equation}
\begin{equation}
\nonumber
  g(\boldsymbol{x})=
  \begin{pmatrix}   
    0 \\  
    A(\Omega+u(\tau))^2\cos\tau-cx_2  
  \end{pmatrix}
  .
\end{equation}

The Melnikov function for system (2) is:
\begin{align}%
    M(\tau) &=\int_{-\infty}^{+\infty} f(q^0_+(\tau)) \wedge g(q^0_+(\tau),\tau+t_0) \,dt  \nonumber\\
    & =\int_{-\infty}^{+\infty} \dfrac{x^0_2(\tau)}{\Omega}[A(\Omega+u(\tau+t_0))^2\cos(\tau+t_0)  \nonumber\\
    & -cx^0_2(\tau)] \,d\tau
    .
\end{align}%
Following the mathematical analysis given in the Appendix, we have:

\begin{align}%
  M(t_0) &  =A(2\pi\Omega^2+4\Omega\kappa\Delta+2\kappa^2\Delta)\times\Phi(\Omega)\sin(t_0) \nonumber\\
  &  +2A\sum_{n=1}^N(2\Omega\kappa+\kappa^2)\times\psi(n)   \nonumber\\
  &  -4c[\dfrac{\ln(\sqrt{\alpha^2+1}-\alpha)}{\sqrt{\alpha^2+1}}+\alpha].
\end{align}
where
\begin{equation}
  \Phi(\Omega)=\sin [\dfrac{\Omega}{\alpha}\sinh^{-1}(\alpha)]\times\sech (\dfrac{\Omega\pi}{2\alpha}),
\end{equation}
\begin{align}%
  \psi(n) &=\dfrac{\sin(n\Delta)}{n}(n-1)\sin \dfrac{\Omega}{\alpha}\sinh^{-1}(\alpha) \nonumber\\
  & \times\sech (\dfrac{\Omega\pi}{2\alpha}(1-n)\sin(t_0-nt_0) \nonumber\\
  & -\dfrac{\sin(n\Delta)}{n}(1+n)\sin \dfrac{\Omega}{\alpha}\sinh^{-1}(\alpha)  \nonumber\\
  & \times\sech (\dfrac{\Omega\pi}{2\alpha}(1+n))\sin(t_0+nt_0),
\end{align}

\begin{figure*}
\centering
\subfigure{
  \includegraphics[scale=0.5]{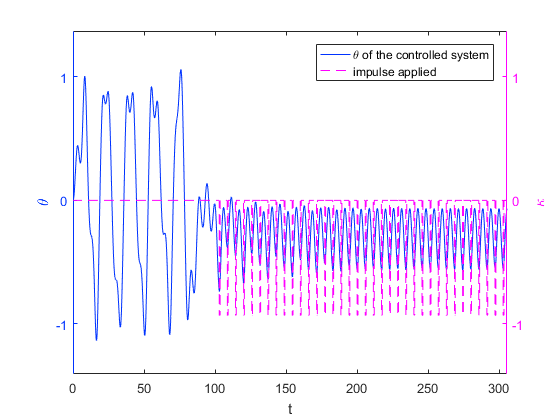}
}
\subfigure{
  \includegraphics[scale=0.5]{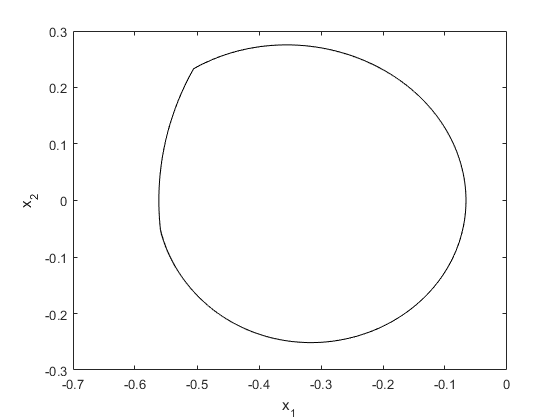}
}
\caption{(a) The state variable $x_1$ is plotted with parameters $\Delta=0.57$ and $\kappa=-0.93$ using blue solid line, magenta dashed line indicates the impulse applied; (b) The corresponding steady state phase portrait after impulse control is activated.}
\end{figure*}

\begin{figure*}
\centering
\subfigure{
  \includegraphics[scale=0.5]{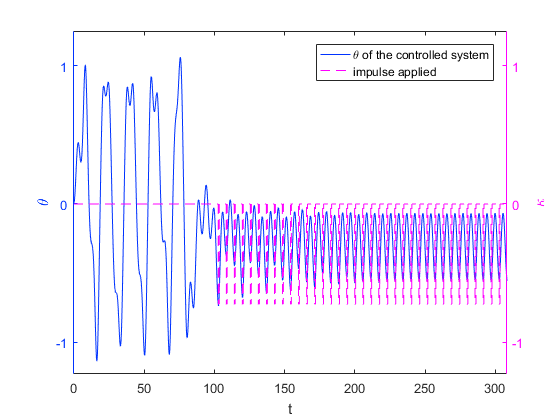}
}
\subfigure{
  \includegraphics[scale=0.5]{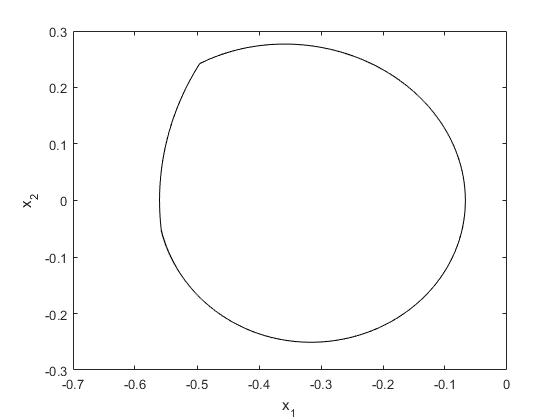}
}
\caption{(a) The state variable $x_1$ is plotted with parameters $\Delta=0.63$ and $\kappa=-0.724$ using blue solid line, magenta dashed line indicates the impulse applied; (b) The corresponding steady state phase portrait after impulse control is activated.}
\end{figure*}
According the theory in [30], the necessary condition for homoclinic chaos is that $M(t_0)$ is equal to zero. This provides a design criterion for chaos suppression, i.e., if $M(t_0)<0$ is satisfied for all $t_0$, then no chaos exists.

Since $\Delta$ is small, $\sin(n\Delta)/n\approx \Delta$, $|\sin(t_0)|<1$, we obtain:
\begin{align}%
  M(t_0) &  <A(2\pi\Omega^2+4\Omega\kappa\Delta+2\kappa^2\Delta)\times\Phi(\Omega) \nonumber\\
  &  +2A\Delta\times(2\Omega\kappa+\kappa^2)\Psi(n)   \nonumber\\
  &  -4c[\dfrac{\ln(\sqrt{\alpha^2+1}-\alpha)}{\sqrt{\alpha^2+1}}+\alpha]<0,
\end{align}
where
\begin{align}%
  \Psi(n) &=\sum_{n=1}^N[(n-1)\sin \dfrac{\Omega}{\alpha}\sinh^{-1}(\alpha)\times\sech (\dfrac{\Omega\pi}{2\alpha}(n-1) \nonumber\\
  & -(1+n)\sin \dfrac{\Omega}{\alpha}\sinh^{-1}(\alpha)\times\sech (\dfrac{\Omega\pi}{2\alpha}(1+n)].  \nonumber\\
\end{align}
Substituting (13) and (14) into (12), we finally get that if:
\begin{equation}
  \Delta>\dfrac{\dfrac{2c}{A}\times(\dfrac{\ln(\sqrt{\alpha^2-1}-\alpha)}{\sqrt{\alpha^2-1}}+\alpha)-\pi\Omega^2\Phi(\Omega)}{(2\Omega\kappa+\kappa^2)(\Phi(\Omega)+\Psi(n))}
\end{equation}
then $M(t_0)<0$ for all $t_0$, and chaos is eliminated. The theorem is proved.

The relationship between $\kappa$ and $\Delta$ for chaos suppression is given in Fig. 4, where the system parameters $\Omega=1.1, c=0.1$, and $A=0.2$. Chaos is quenched in the controlled system if the pairs $(\Delta, \kappa)$ are above the curve. The theorem gives us the parameter range to control chaos in FSRL system.

Simulations are carried out to confirm the validity of the criterion given by inequality (15). Choosing the parameter pair $(\Delta,\kappa)=(0.57,-0.93)$ and $(\Delta,\kappa)=(0.63,-0.724)$, when the proposed impulse control is activated at 100s, the simulation results are given in Figs. 5 and 6. Figures 5a and 6a show the state variable $x_1$, and the corresponding applied impulse. It can be seen that $x_1$ is stabilized resulting in a periodic state after a transient state after the controller is applied. Figures 5b and 6b depict the corresponding steady state phase portraits, respectively.

\section{Conclusions}

Chaos in the FSRL system causes defects in the silicon crystal growth quality, which need to be avoided. Therefore in this paper, the Melnikov method is used to suppress the chaotic dynamical behaviour. An impulse control is proposed to eliminate chaos. We also derive the parameter selection rule for impulse control to suppress chaos using the Melnikov method. We have found that the impulse control is simple, not requiring state feedback, and it is an effective control technique to control chaos in the FSRL system.


%
% ...


\section*{Appendix}
\setcounter{equation}{0}
\renewcommand{\theequation}{A\arabic{equation}}%
In this appendix, the calculation of the Melnikov¡¯s function (10) is carried out. Recalling (9) as follows:
\begin{equation}
\begin{aligned}%
    M(\tau) & =\int_{-\infty}^{+\infty} \dfrac{x_2(\tau)}{\Omega}[A(\Omega+u(\tau+t_0))^2\cos(\tau+t_0) \nonumber\\
    & -cx_2(\tau)] \,d\tau  \nonumber\\
    & =\int_{-\infty}^{+\infty} \dfrac{1}{\Omega}\times\dfrac{-2\alpha^2\sinh(\alpha \tau	\setminus\Omega)}{\alpha^2+\cosh^2(\alpha\tau\setminus\Omega)}[A(\Omega+u(\tau+t_0))^2             \nonumber\\
    & \times\cos(\tau+t_0)-c\dfrac{-2\alpha^2\sinh(\alpha \tau	\setminus\Omega)}{\alpha^2+\cosh^2(\alpha\tau\setminus\Omega)}] \,d\tau\nonumber\\
\end{aligned}%
\end{equation}
$P(t)=(\Omega+u(t))^2$ is also a periodic signal and it can be represented in terms of the Fourier series $P_N(t)$,

\begin{align}%
\nonumber
  P_N(t) & =(\Omega^2+\dfrac{2\Omega\kappa\Delta}{\pi}+\dfrac{\kappa^2\Delta}{\pi})  \nonumber\\
  &  +\sum_{n=1}^N (\dfrac{4\Omega\kappa}{n\pi}+\dfrac{2\kappa^2}{n\pi})\sin({n\Delta})\cos({n\tau}) \nonumber\\
\end{align}

As indicated in [29], the two homoclinic orbits for the unperturbed system of (1) are already obtained,
\begin{equation}
\nonumber
  q^0_+(t)=(2\cot^{-1}(\dfrac{1}{\alpha})\cosh\alpha t,-\dfrac{2\alpha^2\sinh\alpha t}{\alpha^2+\cosh^2\alpha t}),
  \label{Eqa10}
\end{equation}
and
\begin{equation}
\nonumber
  q^0_-(t)=(-2\cot^{-1}(\dfrac{1}{\alpha})\cosh\alpha t,\dfrac{2\alpha^2\sinh\alpha t}{\alpha^2+\cosh^2\alpha t}).
  \label{Eqa11}
\end{equation}
with $q^0_+(t)$ , (9) becomes (A2)(in next page),\\
where
\begin{equation}
  \Phi(\Omega)=\sin [\dfrac{\Omega}{\alpha}\sinh^{-1}(\alpha)]\times\sech (\dfrac{\Omega\pi}{2\alpha}), \nonumber
\end{equation}
and

\begin{align}
\nonumber\\
  \psi(n) &=\dfrac{\sin(n\Delta)}{n}(n-1)\sin \dfrac{\Omega}{\alpha}\sinh^{-1}(\alpha) \nonumber\\
  & \times\sech (\dfrac{\Omega\pi}{2\alpha}(1-n)\sin(t_0-nt_0) \nonumber\\
  & -\dfrac{\sin(n\Delta)}{n}(1+n)\sin \dfrac{\Omega}{\alpha}\sinh^{-1}(\alpha)  \nonumber\\
  & \times\sech (\dfrac{\Omega\pi}{2\alpha}(1+n))\sin(t_0+nt_0),  \nonumber
\end{align}

By this way, we can derive Eq. (10).

The derivation for $q^0_-(t)$ can be derived in the same way.

\begin{figure*}[ht]
\begin{align}%
  M(t_0) &  =\int_{-\infty}^{+\infty} \dfrac{1}{\Omega}\times\dfrac{-2\alpha^2\sinh(\alpha
  \tau\setminus\Omega)}{\alpha^2+\cosh^2(\alpha
  \tau\setminus\Omega)}[A(\Omega^2+\dfrac{2\Omega\kappa\Delta}{\pi}+\dfrac{\kappa^2\Delta}{\pi})+\sum_{n=1}^N (\dfrac{4\Omega\kappa}{n\pi}+\dfrac{2\kappa^2}{n\pi})\sin({n\Delta})\cos({n\tau}))\cos(\tau+t_0)  \nonumber\\
  &  -c\dfrac{-2\alpha^2\sinh(\alpha\tau\setminus\Omega)}{\alpha^2+\cosh^2(\alpha\tau\setminus\Omega)}] \,d\tau \nonumber\\
  &  =A(\Omega^2+\dfrac{2\Omega\kappa\Delta}{\pi}+\dfrac{\kappa^2\Delta}{\pi})\int_{-\infty}^{+\infty} \dfrac{1}{\Omega}\times\dfrac{-2\alpha^2\sinh(\alpha \tau\setminus\Omega)}{\alpha^2+\cosh^2(\alpha\tau\setminus\Omega)}\cos(\tau+t_0)   \,d\tau\nonumber\\
  &  +A\sum_{n=1}^N(\dfrac{4\Omega\kappa}{n\pi}+\dfrac{2\kappa^2}{n\pi})\sin({n\Delta}) \int_{-\infty}^{+\infty} \dfrac{1}{\Omega}\times\dfrac{-2\alpha^2\sinh(\alpha \tau\setminus\Omega)}{\alpha^2+\cosh^2(\alpha\tau\setminus\Omega)}\cos(n\tau)\cos(\tau+t_0)   \,d\tau\nonumber\\
  &  -c\int_{-\infty}^{+\infty}\dfrac{1}{\Omega}\times(\dfrac{-2\alpha^2\sinh(\alpha \tau\setminus\Omega)}{\alpha^2+\cosh^2(\alpha\tau\setminus\Omega)})^2 \,d\tau \nonumber\\
  &  =A(\Omega^2+\dfrac{2\Omega\kappa\Delta}{\pi}+\dfrac{\kappa^2\Delta}{\pi})\times 2\pi\sin(\dfrac{\Omega}{\alpha}sinh^{-1}(\alpha))sech(\dfrac{\pi\Omega}{2\alpha})sin(t_0)    \nonumber\\
  &  +A\sum_{n=1}^N(\dfrac{4\Omega\kappa}{n\pi}+\dfrac{2\kappa^2}{n\pi})\sin({n\Delta}) \int_{-\infty}^{+\infty} \dfrac{1}{\Omega}\times\dfrac{-2\alpha^2\sinh(\alpha \tau\setminus\Omega)}{\alpha^2+\cosh^2(\alpha\tau\setminus\Omega)}[cos(n(\tau+t_0)+(\tau+t_0))   \nonumber\\
  & +cos(n(\tau+t_0)-(\tau+t_0))]\,d\tau\nonumber\\
  & -4c(\dfrac{\ln(\sqrt{\alpha^2-1}-\alpha)}{\sqrt{\alpha^2-1}}+\alpha) \nonumber\\
  &  =A(\Omega^2+\dfrac{2\Omega\kappa\Delta}{\pi}+\dfrac{\kappa^2\Delta}{\pi})\times 2\pi\sin(\dfrac{\Omega}{\alpha}sinh^{-1}(\alpha))sech(\dfrac{\pi\Omega}{2\alpha})sin(t_0)    \nonumber\\
  &  +A\sum_{n=1}^N(\dfrac{4\Omega\kappa}{n\pi}+\dfrac{2\kappa^2}{n\pi})\sin({n\Delta})\time2\pi(n-1) \sin(\dfrac{\Omega}{\alpha}sinh^{-1}(\alpha))sech(\dfrac{\pi\Omega}{2\alpha}(n-1))sin((n-1)t_0)  \nonumber\\
  &  -A\sum_{n=1}^N(\dfrac{4\Omega\kappa}{n\pi}+\dfrac{2\kappa^2}{n\pi})\sin({n\Delta})\time2\pi(n+1) \sin(\dfrac{\Omega}{\alpha}sinh^{-1}(\alpha))sech(\dfrac{\pi\Omega}{2\alpha}(n+1))sin((n+1)t_0)  \nonumber\\
  & -4c(\dfrac{\ln(\sqrt{\alpha^2-1}-\alpha)}{\sqrt{\alpha^2-1}}+\alpha) \nonumber\\
  & =A(2\pi\Omega^2+4\Omega\kappa\Delta+2\kappa^2\Delta)\times\Phi(\Omega)\sin(t_0)+2A\sum_{n=1}^N(2 \Omega\kappa+\kappa^2)\times\psi(n)-4c[\dfrac{\ln(\sqrt{\alpha^2+1}-\alpha)}{\sqrt{\alpha^2+1}}+\alpha].
\end{align}
\end{figure*}

\end{document}